\documentclass[reprint,superscriptaddress,amsmath,amssymb,aps,onecolumn]{revtex4-1}

\usepackage{graphicx}
\usepackage{subfigure}
\usepackage{amssymb}
\usepackage{amsmath} 
\usepackage{bm}
\usepackage{hyperref}
\usepackage{setspace}

\begin{document}

\title{High-Resolution Numerical Simulations of a Large-Scale Helium Plume Using Adaptive Mesh Refinement}

\author{Nicholas T.\ Wimer}
\email{Nicholas.Wimer@Colorado.edu}
\affiliation{University of Colorado Boulder, Boulder, CO}
\author{Marcus S.\ Day}
\affiliation{Lawrence Berkeley National Laboratory, Berkeley, CA}
\author{Caelan Lapointe}
\affiliation{University of Colorado Boulder, Boulder, CO}
\author{Amanda S.\ Makowiecki}
\affiliation{University of Colorado Boulder, Boulder, CO}
\author{Jeffrey F.\ Glusman}
\affiliation{University of Colorado Boulder, Boulder, CO}
\author{John W.\ Daily}
\affiliation{University of Colorado Boulder, Boulder, CO}
\author{Gregory B.\ Rieker}
\affiliation{University of Colorado Boulder, Boulder, CO}
\author{Peter E.\ Hamlington}
\affiliation{University of Colorado Boulder, Boulder, CO}

\date{\today}

%%%%%%%%%%%%%%%%%%%%%%%%%%%%%%%%%%
\begin{abstract}
The physical characteristics and evolution of a large-scale helium plume are examined through a series of numerical simulations with increasing physical resolution using adaptive mesh refinement (AMR). The five simulations each model a 1~m diameter circular helium plume exiting into a (4~m)$^3$ domain, and differ solely with respect to the smallest scales resolved using the AMR, spanning resolutions from 15.6~mm down to 0.976~mm. As the physical resolution becomes finer, the helium-air shear layer and subsequent Kelvin-Helmholtz instability are better resolved, leading to a shift in the observed plume structure and dynamics. In particular, a critical resolution is found between 3.91~mm and 1.95~mm, below which the mean statistics and frequency content of the plume are altered by the development of a Rayleigh-Taylor instability near the centerline in close proximity to the base of the plume. This shift corresponds to a plume ``puffing'' frequency that is slightly higher than would be predicted using empirical relationships developed for buoyant jets. Ultimately, the high-fidelity simulations performed here are intended as a new validation dataset for the development of subgrid-scale models used in large eddy simulations of real-world buoyancy-driven flows.
\end{abstract}
%%%%%%%%%%%%%%%%%%%%%%%%%%%%%%%%%%

\maketitle

%%%%%%%%%%%%%%%%%%%%%%%%%%%%%%%%%%
\section{Introduction\label{sec:intro}}
%%%%%%%%%%%%%%%%%%%%%%%%%%%%%%%%%%

%Motivation
Buoyancy-driven flows are ubiquitous in nature and engineering, spanning natural phenomena as diverse as thermals in the ocean, volcanic plumes, and wildland fires, as well as engineering applications such as flue gas systems and burners for industrial processing. With the increasing availability and decreasing cost of high performance computing resources, numerical simulations are being used more and more to study these flows, in particular using large eddy (e.g., \cite{Bastiaans2000,Zhou2001,Zhou2001a,DesJardin2004,Worthy2005,Pham2007,Burton2009,Blanquart2009,Maragkos2012,Maragkos2013,Jatale2015}) and direct numerical (e.g., \cite{Bastiaans2000,Pham2007}) simulations (LES and DNS, respectively). However, prior attempts to reproduce experimental results for larger-scale buoyancy-driven flows, particularly in the context of LES for meter-scale plumes and pool fires \cite{Brown2018}, have proven difficult, limiting the use of simulations for the understanding and design of many real-world systems. The reasons for this lack of success are still not fully understood, but may be attributable to overly simplistic representations of experimental configurations, inadequate knowledge of experimental initial and boundary conditions, inadequate spatial resolution in simulations, or inaccurate computational models for fluid properties and subgrid-scale (SGS) quantities. In all likelihood, discrepancies between experimental and computational results are a combination of these causes.

%Solution
In the present study, we partially address these discrepancies by performing a series of high-fidelity computational simulations of a large-scale forced helium plume at different grid resolutions. These simulations are used here to examine the structure and dynamics of buoyancy-driven flows, but can also be used for the future development and validation of SGS models. Most notably, the present computations have well-defined initial and boundary conditions, avoid the use of SGS models, and are grid-converged up to second-order statistics. Particular attention is paid to the spatial resolution required to accurately capture the dominant plume dynamics, and we show that the necessary resolution is much finer than that used in any previous computational study of large-scale buoyancy-driven flows.

%=================================
\subsection{Background}
%=================================
%Prior plume studies, experimental and computational
Buoyancy-driven flows have been studied in a wide range of contexts, including pure (i.e., unforced) \cite{Rouse1952,Morton1955,Wakitani1985,Dai1994a,Shabbiri1994,Sreenivas2000,Hunt2001,Fanneløp2003,Whittaker2006,Nam1993,Pham2006,Pham2005,Pham2007} and forced \cite{Morton1958,Mollendorf1973,Papanicolaou1987,Papanicolaou1988,Kyle1993,Panchapakesan1993,Ai2006,Soteriou2002} plumes, where the latter are often referred to as buoyant jets. These studies have placed substantial emphasis on understanding far-field and asymptotic scaling relations, as well as on quantifying the effects of buoyancy on the break-up and stability of jets. The spatio-temporal characteristics of plumes have also received considerable attention, particularly by Cetegen and collaborators, who have examined the characteristic flow oscillations (typically referred to as ``puffing'') that develop in both steadily and unsteadily forced axisymmetric buoyant jets \cite{Cetegen1996,Cetegen1997,Cetegen1997a}, as well as unforced planar plumes \cite{Cetegen1998a}. Other configurations, spanning additional physical phenomena and effects, have revealed further insights into buoyancy-driven flows, including the effects of acoustic noise on vortex evolution in axisymmetric buoyant jets \cite{Becker1968}, the interactions of plumes, jets, and buoyant jets with inclined surfaces \cite{Gebhart1973}, the characteristics of hot air plumes \cite{George1977}, and the interactions between two initially separate round turbulent plumes \cite{Kaye2004}. Buoyancy effects on the structure and dynamics of reacting flows have also been the subject of much research, with an emphasis on determining the characteristics of pool fires \cite{Morton1965,Zukoski1981,Zukoski1984,Weckman1989,Cetegen1993,Lingens1996,McGrattan1996,Desjardin1999,Annarumma1991,Hamins1992,Xin2002,Jiang2003,Zhou2002}. 

%FLAME facility detail and importance, specifically
These prior studies were focused on relatively small-scale flows, with plume diameters no greater than 0.5~m. To better understand large-scale fire dynamics, particle image velocimetry and planar laser induced fluorescence (PLIF) were used at the Fire Laboratory for Accreditation of Models by Experimentation (FLAME) facility \cite{Tieszen2005,OHern2005} at Sandia National Laboratory to study a range of both reacting and non-reacting buoyancy-driven flows. In one such study, velocity and helium mass fraction were measured at several locations above a 1~m diameter, moderately forced, helium plume \cite{OHern2005}. The dynamics and structure of the helium plume bear many similarities to large-scale fires, and this case is thus appealing from a computational perspective because buoyancy-driven aspects of fire dynamics can be studied without the need to account for complex chemical reactions. 

%The various attempts to model the FLAME facility
An early LES of the helium plume in an idealized (i.e., not attempting to reproduce the full geometrical complexity of the experiment) configuration by Desjardin \emph{et al.} \cite{DesJardin2004} was shown to be sensitive to grid resolution, but to have fair agreement with experimental results, despite an over-prediction of the axial turbulence levels and helium concentration near the base of the plume. The inability of standard SGS models to capture up-scale (or backscatter) of energy was also identified as a potential issue. The minimum grid spacing in this study was 1.6 cm, with a maximum spacing of 7.8~cm, resulting in a simulated puffing frequency of 1.5~Hz, which was slightly higher than the frequency of $1.37\pm 0.1$~Hz measured experimentally \cite{OHern2005}. Chung and Devaud \cite{Chung2008} later performed LES of the same idealized configuration using a finer uniform grid resolution of 1.25~cm, resulting in a smaller puffing frequency of 1.4~Hz that was more in line with experiments. Despite generally good agreement with experimental measurements of velocity, predicted helium mass fraction measurements deviated from the experiments, particularly near the centerline. Similarly, Blanquart \emph{et al.} \cite{Blanquart2009} were able to obtain good agreement with experimental velocity measurements by modeling the full geometry of the FLAME facility using LES, but discrepancies with the experiments remained for measurements of the helium mass fraction. Burton \cite{Burton2009} attempted to address the SGS modeling deficiencies first identified by Desjardin \emph{et al.} \cite{DesJardin2004} through the use of the nonlinear-LES method, which permits backscatter of energy from small to large scales, but discrepancies between the computational and experimental results remained. Maragkos \emph{et al.} \cite{Maragkos2012,Maragkos2013} were able to obtain reasonable agreement between computational and experimental axial velocities using LES with the Smagorinsky SGS model, but the helium mass fraction was significantly over-predicted at all heights, which was attributed to the lack of differential diffusion in the simulations. The minimum grid spacing used by Maragkos \emph{et al.} \cite{Maragkos2013} was 1.23~cm, with a maximum spacing of 5.39~cm, resulting in a puffing frequency of 1.41~Hz with an SGS model and 1.31~Hz without an SGS model. More recently, Jatale \emph{et al.} \cite{Jatale2015} presented and applied a framework for uncertainty quantification to the 1~m diameter helium plume using the experimental data and LES. 

%Continuing problem
To summarize these prior studies, no approach has been completely successful in capturing both velocity and helium mass fraction experimental results \cite{Brown2018}. In particular, improvements in grid resolution down to roughly 1~cm, the use of different SGS models, including those that permit backscatter, and more accurate representations of the experimental geometry have all failed to provide good agreement between simulations and the experiments. Perhaps of even greater concern is the possibility that, given the results from Blanquart \emph{et al.} \cite{Blanquart2009} where the geometry of the FLAME facility was modeled, there is insufficient information regarding experimental boundary data and material properties to fully characterize the experimental conditions. 

Given this situation, there is a need for a new large-scale helium plume dataset that is similar to the previous plume experiment at the FLAME facility, but uses exactly determined boundary conditions and material properties. This can be accomplished through the use of a high-fidelity numerical simulation without an SGS model that is converged up to, at least, second-order statistics. With this high-fidelity computational dataset, the structure and dynamics of large-scale buoyancy-driven flows can be examined and the accuracy of different SGS models can be determined, with no ambiguity as to the correct initial and boundary conditions, or material properties.

%=================================
\subsection{Present approach}
%=================================

%Give overview of current approach
In the present study, we examine the structure and dynamics of a 1~m helium plume through a series of numerical simulations with no SGS model. The physical resolution of the simulations is successively refined to obtain fully converged velocity and helium mass fraction statistics up to second order. The simulations leverage the power of adaptive mesh refinement (AMR) to locally refine the grid in regions of high density gradient and vorticity. Using AMR, we are able to resolve the 1~m helium plume from a base resolution of 31.25~mm down to sub-millimeter scales in regions of interest. We show that the simulations are converged up to second-order velocity and helium mass fraction statistics at a resolution of roughly 2~mm. This indicates that the present simulations can be considered equivalent to DNS for these statistics, although higher-order statistics are likely to remain only marginally converged at even the finest resolutions used here. Similarly, we show that the puffing frequency is approximately captured at a resolution of 2~mm, with a frequency that is slightly greater than that predicted based on prior experimental studies of axisymmetric buoyancy-driven flows \cite{Cetegen1996,Cetegen1997,Cetegen1997a}.

%Caveats of the present study
It should be noted that, in the present study, we do not attempt to quantitatively match the results of the Sandia helium plume experiment. As noted in the previous section, the geometrical complexity of the FLAME facility is substantial, and there is also incomplete information on the boundary conditions present during the experiments. As such, the configuration studied here is related to, but not the same as, the Sandia helium plume experiment. The focus here is thus on providing a new dataset that is fully converged up to second-order statistics, for which initial and boundary conditions, as well as all material properties, are exactly known. With this dataset, SGS models can be developed and tested with no uncertainty regarding the validation data.

%Outline current paper
The outline of the paper is as follows. A description of the numerical simulations is provided in the next section, including details of the numerical method and the physical configuration modeled. Section \ref{sec:results} describes and discusses results from the simulations, with a particular focus on first- and second-order statistics, as well as the puffing frequency of the plume. Finally, a summary and conclusions are provided at the end.

%%%%%%%%%%%%%%%%%%%%%%%%%%%%%%%%%%
\section{Description of Numerical Simulations}
%%%%%%%%%%%%%%%%%%%%%%%%%%%%%%%%%%
The numerical code used to perform the numerical simulations is \texttt{PeleLM}, a low-Mach, adaptive, parallel, reacting flow code developed at Lawrence Berkeley National Laboratory (LBNL) \cite{Almgren1998,Day2000,Bell2005,Nonaka2012,Nonaka2018}. Details of the equations solved in the present simulations, the adaptive numerical method, and the physical and computational configurations of the simulations are outlined in the following. 

%=================================
\subsection{Low-Mach governing equations}
%=================================
The simulations solve the fully compressible momentum (i.e., Navier Stokes) equation, including a gravitational body force, given by \cite{Nonaka2018}
\begin{equation}\label{eq:ns}
\frac{\partial u_i}{\partial t} +u_j\frac{\partial u_i}{\partial x_j} = - \frac{1}{\rho} \frac{\partial \pi}{\partial x_i}+ g_i+\frac{1}{\rho} \frac{\partial}{\partial x_j}\left[\mu\left(\frac{\partial u_i}{\partial x_j}+\frac{\partial u_j}{\partial x_i} - \frac{2}{3}\delta_{ij} \frac{\partial u_k}{\partial x_k}\right)\right]\,,
\end{equation}
where $u_i$ is the velocity, $\rho$ is the density, $g_i$ is the gravitational acceleration, and $\mu$ is the viscosity. The pressure perturbation field $\pi(\bm{x},t)$ in Eq.~\eqref{eq:ns} is superimposed on a homogeneous and steady background component, $p_0$, and thus the total pressure $p(\bm{x},t)$ is written as 
\begin{equation}
	p(\bm{x},t) = p_0 + \pi (\bm{x},t) \,.
	\label{eq:pertpress}
\end{equation}
Assuming that $\pi/p_0$ is $\mathcal{O}(M^2)$ and that the Mach number is small, $\pi$ can be considered as a correction to the pressure necessary to ensure that $p_0$ remains homogeneous \cite{Nonaka2012,Nonaka2018}. The background state $p_0$ is assumed to be thermodynamically consistent with $\rho$ and the temperature $T$ according to the ideal gas equation of state
\begin{equation}
	p_0 = \rho \mathcal{R} T \sum_\alpha \frac{Y_\alpha}{W_\alpha} \,,
	\label{eq:eos}
\end{equation}
where $\mathcal{R}$ is the universal gas constant, and $Y_\alpha$ and $W_\alpha$ are the mass fraction and molecular weight of species $\alpha$, respectively. The perturbation pressure $\pi$ is determined by a divergence constraint on the velocity field, which will be discussed later in this section.

The dynamics of $Y_\alpha$ and the sensible enthalpy, $h$, are described by the non-reactive transport equations
\begin{align}
	\frac{\partial(\rho Y_\alpha)}{\partial t} + \frac{\partial (u_j\rho Y_\alpha)}{\partial x_j} &= \frac{\partial \Gamma_{\alpha j}}{\partial x_j} \,, \label{eq:rhoYm} \\
	\frac{\partial(\rho h)}{\partial t} + \frac{\partial (u_j\rho h)}{\partial x_j} &= \frac{\partial}{\partial x_j}\left( \lambda \frac{\partial T}{\partial x_j}\right) + \frac{\partial}{\partial x_j}\left(\sum_\alpha h_\alpha \Gamma_{\alpha j}\right) \label{eq:rhoh} \,,
\end{align}
where the diffusive flux $\Gamma_{\alpha j}$ is given by
\begin{equation}
    \Gamma_{\alpha j} = \rho D_\alpha \frac{W_\mathrm{mix}}{W_\alpha} \frac{\partial Y_\alpha}{\partial x_j}\,.
\end{equation}
In the above equations, $D_\alpha(T)$ and $h_\alpha(T)$ are, respectively, the temperature-dependent diffusivity and enthalpy of species $\alpha$, $\lambda$ is the thermal diffusivity, and summation over repeated Greek indices is not implied. The sensible enthalpy is expressed as $h=\sum_\alpha Y_\alpha h_\alpha$. The mixture average molecular weight, $W_\mathrm{mix}$, appearing in Eqs.~\eqref{eq:rhoYm} and \eqref{eq:rhoh} is assumed to be weakly varying spatially, and is given by $W_\mathrm{mix} = (\sum_\alpha Y_\alpha/W_\alpha)^{-1}$. The viscosity $\mu(T,Y_\alpha)$ in Eq.~\eqref{eq:ns} and the thermal diffusivity $\lambda(T,Y_\alpha)$ in Eq.~\eqref{eq:rhoh} are given by thermodynamic relations that depend on $T$ and the fluid composition expressed using $Y_\alpha$. 

It should be noted that pressure does not appear explicitly in Eq.~\eqref{eq:rhoh} because $p_0$, which is thermodynamically consistent with $h$ and the other thermodynamic variables, is a constant. Soret and Dufour effects are also neglected in the present simulations, and changes in $h$ due to viscous dissipation are assumed to be negligible.

By summing Eq.~\eqref{eq:rhoYm} over all species (i.e., all $\alpha$) and noting that $\sum_\alpha \Gamma_{\alpha j} = 0$ (since species diffusion can neither create nor destroy mass), we obtain the continuity equation expressing conservation of mass, namely
\begin{equation}
\frac{\partial \rho}{\partial t} + \frac{\partial (u_j \rho)}{\partial x_j} = 0 \,.
\label{eq:continuity}
\end{equation}
Following Nonaka \emph{et al.} \cite{Nonaka2018}, this equation can be re-expressed using the equation of state for $p_0$ in Eq.~\eqref{eq:eos} and the transport equations for $Y_\alpha$ and $h$ in Eqs.~\eqref{eq:rhoYm} and \eqref{eq:rhoh} to give the divergence constraint 
\begin{equation}
 \frac{\partial u_i}{\partial x_i}= \frac{1}{\rho c_p T} \left[\frac{\partial}{\partial x_j}\left(\lambda \frac{\partial T}{\partial x_j}\right) + \frac{\partial}{\partial x_j}\left(\sum_\alpha h_\alpha \Gamma_{\alpha j}\right)\right] + \frac{1}{\rho} \sum_\alpha \left( \frac{W_\mathrm{mix}}{W_\alpha} - \frac{h_\alpha}{c_p T} \right) \frac{\partial \Gamma_{\alpha j}}{\partial x_j} \,,
	\label{eq:divUconstrained}
\end{equation}
where $c_p=c_p(T)$ is the specific heat capacity at constant pressure. In the simulations, we apply an approximate projection to decompose a predicted new-time velocity into an update of $\pi$ and a final new-time velocity that satisfies the divergence constraint given by Eq.\ \eqref{eq:divUconstrained}, as described by Nonaka \emph{et al.} \cite{Nonaka2018}.

The set of partial differential equations given by Eqs.~\eqref{eq:ns}, \eqref{eq:rhoYm}, \eqref{eq:rhoh}, and \eqref{eq:continuity}, combined with thermodynamic relations for the transport coefficients, species enthalpies, and specific heat capacity, constitute the numerical model solved by \texttt{PeleLM}. This equation set permits local compressibility (i.e., variable density) effects due to mass and thermal diffusion, while filtering out acoustic wave propagation. This approach thus enables larger time steps than in a traditional, fully-compressible solver by a factor of $\mathcal{O}(M^{-2})$.

%=================================
\subsection{Numerical approach}
%=================================
The time integration scheme used in \texttt{PeleLM} couples advection and diffusion via iterative re-integration. Each physical process is integrated using time-lagged source terms to account for the other process. For advection, we use a second-order Godunov scheme and for diffusion we use a semi-implicit discretization. The overall numerical method is second-order accurate in both space and time, and the time-step is dynamically determined according to the advective Courant-Friedrichs-Levy (CFL) condition. A more comprehensive description of the numerical approach, algorithm, and methods is available in Ref.~\cite{Nonaka2018}. All transport properties of the fluid (i.e., $\mu$, $D_\alpha$, and $\lambda$) are calculated as mixture-averaged coefficients assuming a mixture of perfect gases.

The adaptive grid approach employed by \texttt{PeleLM} is described in detail by Nonaka \emph{et al.} \cite{Nonaka2018}, and the governing equations outlined in the previous section are solved on a series of uniform, nested grids with no SGS modeling. The solution is first obtained on the base grid (denoted level 0) with cell size $\Delta x$ and time step $\Delta t$. Regions of interest in the domain are tagged for refinement according to user-specified information. New grids are created (i.e., level 1) overlying the tagged region each with cell size $\Delta x/2$. The governing equations are then solved with time step $\Delta t/2$ across level 1 using initial and boundary conditions interpolated from the underlying grid, as necessary. If another level of refinement is added, the process repeats on level 2 using $\Delta x/4$ and $\Delta t/4$; otherwise, level 1 is progressed another $\Delta t/2$ time step to be consistent with the base grid. When grids at adjacent levels in the mesh hierarchy have been advanced to the same physical time, the levels are then synchronized by averaging information from the finest level onto coarser levels, and by adjusting the fluxes along interfaces between coarse and fine levels. The strategy for this adjustment is consistent with that used for the original level advance, while also ensuring discrete conservation across the composite, multi-level time step~\cite{Day2000}.

%=================================
\subsection{Physical configuration}
%=================================
The numerical simulations model the flow above a helium plume with a 1~m diameter and largely follow the idealized setups of previous simulations \cite{Chung2008,DesJardin2004,Maragkos2012,Maragkos2013}, which are themselves based loosely on the experimental configuration from O'Hern \emph{et al.} \cite{OHern2005}. The domain is 4~m $\times$ 4~m $\times$ 4~m, with a 1~m diameter helium inlet that is surrounded by an annular wall of width 0.5~m, included to ensure that entrainment of ambient air occurs primarily in the radial direction. Outside of the 2~m diameter region defined by the plume and annular wall is a slow air co-flow with velocity $V_\mathrm{coflow} = 0.05$~m/s at an atmospheric pressure of 1 bar and a temperature of $T_\mathrm{coflow} = 286$ K. The composition of the air coflow is 79\% nitrogen (N$_2$) and 21\% oxygen (O$_2$). All temperature- and species-dependent relations for $\mu$, $\lambda$, $h_\alpha$, $D_\alpha$, and $c_p$ were obtained from Chemkin-type transport and thermodynamic files that depend on the gas phase temperature, $T$, and composition, $Y_\alpha$.

At the beginning of the simulation, the domain was filled with quiescent air at 286~K and 1~bar. The helium mixture entered the domain with a velocity of $V_\mathrm{plume} =  0.325$~m/s and temperature $T_\mathrm{plume} =  284$~K. Consistent with previous simulations \cite{Burton2009,Blanquart2009,Chung2008,DesJardin2004,Maragkos2012,Maragkos2013}, the plume was not pure helium, but rather a mixture of 96.4\% helium (He), 1.9\% oxygen (O$_2$), and 1.7\% acetone (CH$_3$COCH$_3$). The resulting mean molecular weight of the helium plume was approximately 5.45~g/mol. We preserved the presence of acetone in the present study for future experimental work that may require its presence as a fluorescent tracer for PLIF data. Similarly, contrary to previous attempts to model the Sandia FLAME facility, which is at higher altitude and has an ambient pressure of roughly 0.8~bar, we used a 1~bar ambient pressure to create a more generic and widely applicable dataset. It should also be noted that the ambient pressure plays no role in determining the non-dimensional parameters that characterize the buoyancy force, such as the Richardson number which depends only on the pressure-independent ratio of the plume source and ambient air densities.

The bottom of the domain, composed of the helium plume, the wall pedestal region, and the air-coflow, was specified using Dirichlet boundary conditions. The top and four sides of the domain were open boundary conditions with velocity fields computed to satisfy the divergence constraint in Eq.~\eqref{eq:divUconstrained}, allowing for entrainment of ambient air or outflow of helium-air mixture as needed. 

%=================================
\subsection{Simulations performed}
%=================================
The simulations presented here were all initialized on the same level 0 base grid with 128 $\times$ 128 $\times$ 128 cells in a 4~m $\times$ 4~m $\times$ 4~m domain, resulting in a uniform base resolution of 31.25~mm, using dynamic time-stepping governed by a maximum CFL condition of 0.7. The simulations were allowed to develop on this coarse grid for 30~s to remove any transients resulting from the initial conditions. After the first 30~s, varying levels of AMR were used to resolve the flow down to smaller and smaller length scales. After the addition of the AMR, the simulations were then run for another 30~s. The first 10~s of this restart period were ignored in the statistical analysis to ensure that the AMR procedure had enough time to adapt to the flow field. Consequently, all time-averaged statistics presented in Section~\ref{sec:results} were computed over a 20~s period.

%---------------------------------
\begin{table}[t!]
\centering
\singlespacing{\caption{\label{tab:sims} Details of the numerical simulations (denoted R1--R5) performed in the present study, including the number of AMR levels in each simulation, the corresponding resolution of the finest grid cells, the number of computational grid cells at the finest level, the total number of grid cells, and the local time step at the finest AMR level. The percentage volume fractions covered by the cells at the finest AMR levels are also indicated. The base grid (corresponding to AMR level 0) in each simulation was $128\times 128 \times 128$ and the physical domain size was 4~m $\times$ 4~m $\times$ 4~m.}}
\begin{tabular}{cccccc}
\hline
\hline
Simulation  & AMR Levels    & Resolution    & Cells at finest level                 & Total cells       & Local time step \\
\hline
R1          & 1         & 15.62 mm      & $1.5\times10^6$ (9.2\% of domain)    & $3.6\times10^6$   & $1.5\times10^{-3}$ s \\
R2          & 2         & 7.81 mm       & $3.3\times10^6$ (2.5\% of domain)    & $6.9\times10^6$   & $7.5\times10^{-4}$ s \\
R3          & 3         & 3.91 mm       & $11\times10^6$  (1.0\% of domain)    & $18\times10^6$    & $3.8\times10^{-4}$ s \\
R4          & 4         & 1.95 mm       & $32\times10^6$  (0.37\% of domain)   & $50\times10^6$    & $1.9\times10^{-4}$ s \\
R5          & 5         & 0.976 mm       & $15\times10^6$  (0.02\% of domain)   & $65\times10^6$    & $9.4\times10^{-5}$ s \\
\hline
\hline
\end{tabular}
\end{table}
%---------------------------------

Five separate simulations were performed, each with identical initial and boundary conditions; the only difference between the simulations was the number of AMR levels used. Each additional level of AMR resulted in a halving of the smallest cell size, thereby increasing the resolution. The base grid (level 0) had a uniform cell size of 31.25~mm, and therefore the resolved scale of the simulation with one level of AMR was 15.62~mm, two levels of AMR resolved 7.81~mm, three levels resolved 3.91~mm, four levels resolved 1.95~mm, and finally the simulation with five levels of AMR resolved from a base grid of 31.25 mm all the way down to 0.976~mm. A summary of the grid cell statistics for each of the five simulations, denoted R1--R5 for increasing numbers of AMR levels, is provided in Table~\ref{tab:sims}.

The grid refinement was determined based on vorticity magnitude and the gradients of density in all three Cartesian directions. At every time step, the refinement criteria was re-evaluated and the domain was re-meshed to ensure that regions of high density gradient or vorticity magnitude were appropriately resolved up to the specified maximum level of refinement and, conversely, that regions of low density gradient or vorticity magnitude were unrefined so that computational resources were conserved. The resulting domains were highly resolved along the interfacial helium-air shear mixing layer and less resolved at the top and sides, which predominantly consist of ambient air or well-mixed helium/air exiting the top of the domain. 

%---------------------------------
\begin{figure}[t!]
\centering
\includegraphics[width=\textwidth]{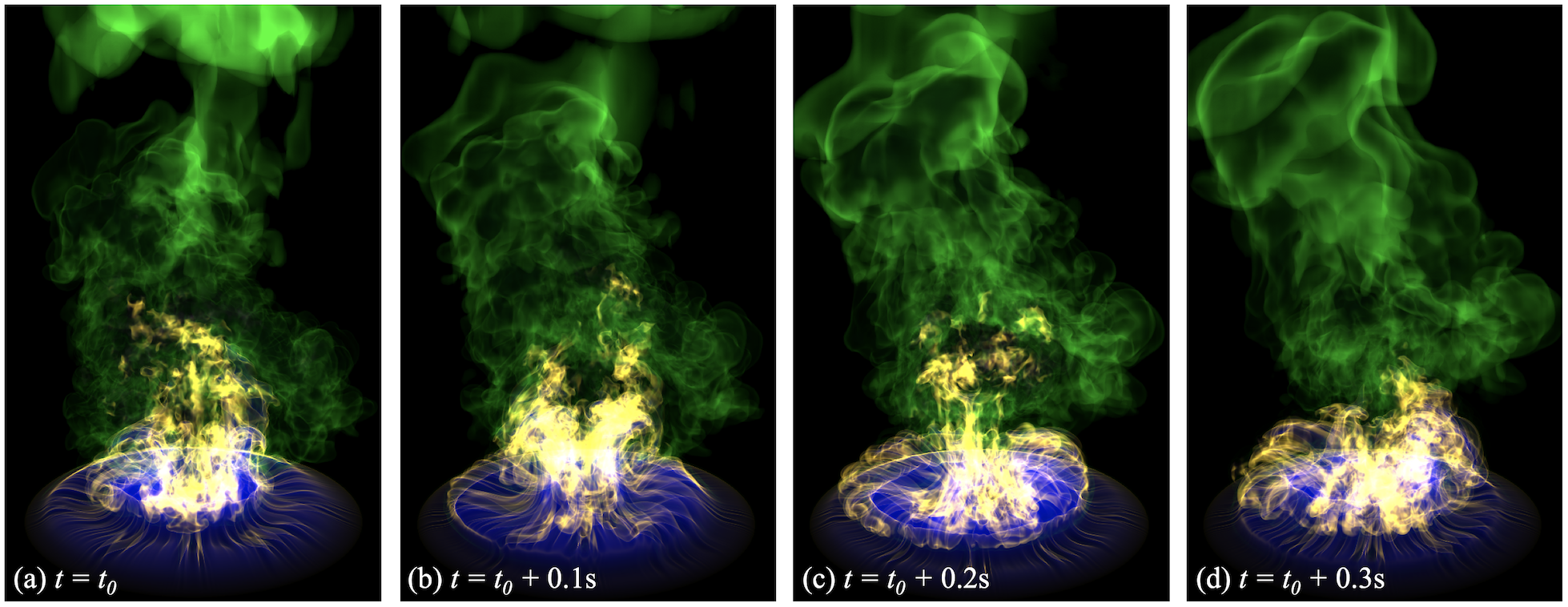}
\singlespacing{\caption{\label{fig:dens} Three-dimensional instantaneous fields of density at four instants (a-d), separated by 0.1~s, during the analysis phase of simulation R4 (see Table \ref{tab:sims}). Lighter colors indicate regions of lower density.}}
\end{figure}
%---------------------------------

Figure~\ref{fig:dens} shows four instantaneous snapshots of three-dimensional density fields from the R4 simulation, revealing the geometric setup of the simulations, as well as the complex physical structures that are resolved. Helium enters the domain from the 1~m diameter inlet, and ambient air is entrained radially. The helium inflow combined with the entrainment of ambient air results in a characteristic ``pooling'' and ``necking'' of the helium that leads to a ``puffing'' of the plume. Further details on the plume structure and dynamics are provided in Section~\ref{sec:results}. An analysis of the puffing frequency, in particular, is provided in Section~\ref{subsec:puffing}.

%%%%%%%%%%%%%%%%%%%%%%%%%%%%%%%%%%
\section{Results and Discussion\label{sec:results}}
%%%%%%%%%%%%%%%%%%%%%%%%%%%%%%%%%%
In the following, instantaneous and statistical simulation results are presented to understand the structure and dynamics of the helium plume, with an emphasis on convergence of the results as the grid resolution is refined. Temporal averages are denoted by $\langle (\cdot) \rangle$ and are computed over a 20~s duration. Temporal and azimuthal averages are denoted by $\overline{(\cdot)}$ and are again computed over a 20~s duration, as well as all azimuthal angles. 

%=================================
\subsection{Qualitative plume dynamics}
\begin{figure}[t!]
\centering
\includegraphics[width=\textwidth]{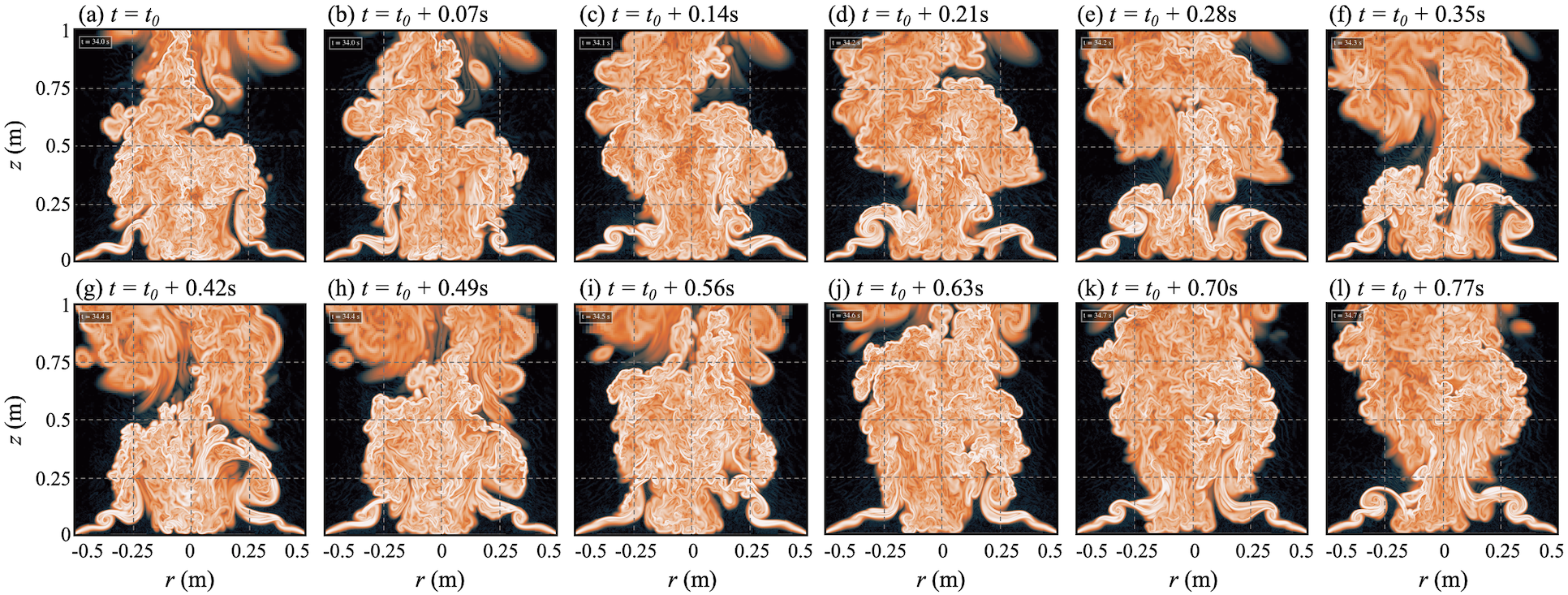}
\singlespacing{\caption{\label{fig:dengrad} Two-dimensional instantaneous fields of the magnitude of the density gradient, $\partial \rho/\partial x_i$, from simulation R5 (see Table \ref{tab:sims}), showing the Kelvin-Helmholtz instability forming along the helium-air shear layer and the Rayleigh-Taylor spike forming a pocket of intense mixing in the center, near the base of the plume. Twelve snapshots are shown, each separated by 0.07~s. Positive $r$ values correspond to an azimuthal angle of $0^\circ$ and negative values correspond to $180^\circ$.}}
\end{figure}

%=================================
Perhaps the most distinctive feature of vertically-oriented buoyant flows is the characteristic puffing motion that results from the interaction of the buoyant force and the shear layer at the edge of the plume. Figure~\ref{fig:dengrad} shows snapshots of the density gradient magnitude in $r-z$ slices at the center of the plume (with positive $r$ values corresponding to an azimuthal angle of $0^\circ$ and negative values corresponding to $180^\circ$). Bright regions in Figure~\ref{fig:dengrad} indicate areas of high density gradient magnitude -- areas in which helium and air are mixed most intensely. These regions of interfacial mixing are key to understanding the complex dynamics of many buoyancy-driven flows. For a non-reacting buoyant plume, these mixing regions lead to a decrease in the buoyant force as the flow entrains ambient air. Conversely, for a reacting pool fire, these are the regions in which fuel and oxidizer mix, resulting in chemical reactions, a local temperature increase, and a corresponding increase in the buoyant force.

%---------------------------------
%---------------------------------

Figure~\ref{fig:dengrad} shows that, over the course of a puffing cycle, the plume structure evolves in several stages. As the helium enters the domain, it is accelerated upwards due to buoyancy, causing it to neck inwards as it moves upwards. The helium drags along ambient air as it necks inwards, creating a helium-air shear layer. This creates a large region around the edge of the plume where helium and air are moving radially inwards at a similar velocity, with the more dense air on top of the less dense helium. As the helium-air shear layer develops, it rolls up via a Kelvin-Helmholtz (KH) instability, as indicated by the sequence of fields in Figures \ref{fig:dengrad}(a-d). The instability travels from the edge of the plume towards the center, grows in size, and eventually pinches off into a toroidal vortex ring, as seen in Figures \ref{fig:dengrad}(f-h). The process of KH roll-up is continuous, and Figures \ref{fig:dengrad}(i-l) show the process repeating again. 

When a coherent vortex ring is shed, it further accelerates the center of the plume through the hole in the toroid, drawing more ambient air deeper into the center of the plume. Some of this helium-air mixture is advected upwards along with the toroidal vortex ring; the remaining helium-air mixture rests atop the highly buoyant helium inlet, now creating a Rayleigh-Taylor (RT) instability. The helium-air mixture in the center of the plume falls down towards the base of the plume in a characteristic RT spike, while the KH roll-up region elevates in an axial RT bubble. The RT instability creates a center region in close proximity to the inlet where helium and air are mixed, resulting in tightly packed, interwoven regions of high density gradient magnitude.

%=================================
\subsection{Time-averaged fields of first- and second-order statistics}
%=================================
The complex dynamics and structure of the large-scale helium plume are a consequence of the interfacial helium-air shear layer, rolling up via a KH instability that is subsequently shed as a vortex ring. Because the plume dynamics are contingent on the development of the KH instability, the numerical simulation results are highly dependent upon physical resolution until the helium-air shear layer is adequately resolved. 

\begin{figure}[t!]
\centering
\includegraphics[scale=1]{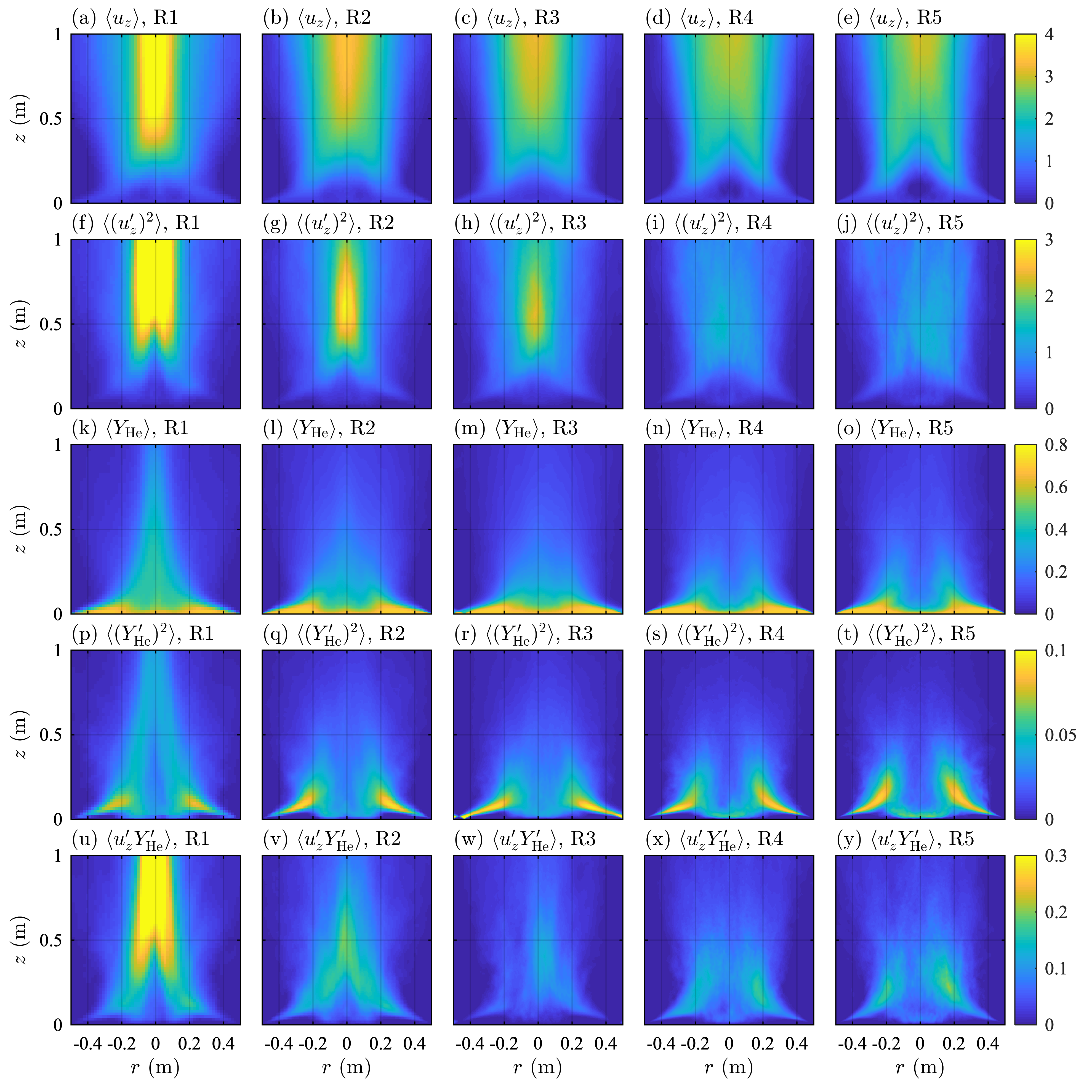}
\singlespacing{\caption{\label{fig:xzaves} Time-averaged $r-z$ slices through the center of the plume of (a-e) mean axial velocity $\langle u_z\rangle$ (in units of m/s), (f-j) axial velocity variance $\langle (u'_z)^2\rangle$ (in units of m$^2$/s$^2$), (k-o) mean helium mass fraction $\langle Y_\mathrm{He}\rangle$, (p-t) helium mass fraction variance $\langle (Y'_\mathrm{He})^2\rangle$, and (u-y) axial flux of helium mass fraction $\langle u'_z Y'_\mathrm{He}\rangle$ (in units of m/s). Columns from left to right show results from simulations with one to five levels of AMR (denoted R1--R5; see Table \ref{tab:sims}). Positive $r$ values correspond to an azimuthal angle of $0^\circ$ and negative values correspond to $180^\circ$.}}
\end{figure}

The effects of resolution are indicated in Figure~\ref{fig:xzaves}, which shows time-averaged slices of axial velocity $\langle u_z\rangle$, axial velocity variance $\langle (u'_z)^2\rangle$, helium mass fraction $\langle Y_\mathrm{He}\rangle$, helium mass fraction variance $\langle (Y'_\mathrm{He})^2\rangle$, and axial helium flux $\langle u'_z Y'_\mathrm{He}\rangle$. As the physical resolution is varied from R1--R5, there are significant changes in the structure and magnitude of all fields.

When the physical resolution is high, as shown in Figure~\ref{fig:xzaves}(o), there is sufficient resolution to fully capture the KH instability and the time-averaged helium mass fraction, namely $\langle Y_\mathrm{He}\rangle$, displays a spike (corresponding to a region of small $\langle Y_\mathrm{He}\rangle$) near the center of the plume close to the inlet as a result of the RT instability. There is a rapid transition to turbulent mixing as suggested by the fluctuating statistics in Figure~\ref{fig:xzaves}, and the variance in helium mass fraction shows the appearance of bubble formation on either side of the RT downward spike as the resolution improves [Figures~\ref{fig:xzaves}(p-t)]. 

%---------------------------------
%---------------------------------

As the physical resolution worsens, the variance in axial velocity becomes lifted off the base of the plume because the early onset of turbulence is suppressed due to a poorly resolved KH instability [Figures~\ref{fig:xzaves}(f-j)]. Additionally, the axial flux of helium mass fraction, $\langle u'_z Y'_\mathrm{He}\rangle$ in Figures~\ref{fig:xzaves}(u-y), begins to lose the structure of the RT instability, as seen by the spike and corresponding bubbles. At the lowest physical resolution, all time-averaged quantities begin to show a clear peak near the center of the domain that is substantially more lifted than in the highest-resolution simulations. Again, this can be attributed to the delayed onset of turbulent mixing due to the poorly resolved helium-air shear layer and resulting instabilities.

%---------------------------------
\begin{figure}[t!]
\includegraphics[scale=1]{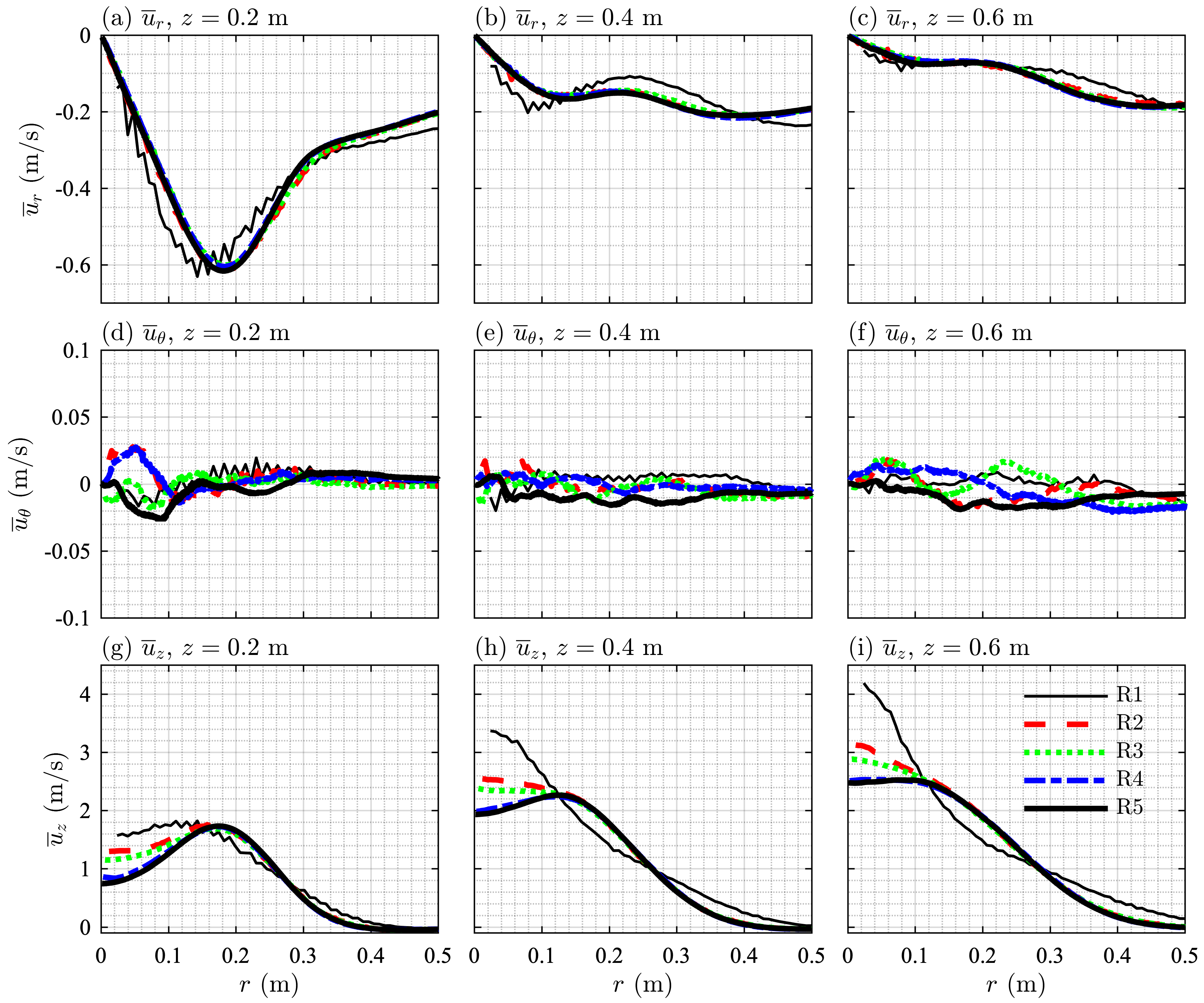}
\singlespacing{\caption{\label{fig:vels} Time and azimuthal averages of (a-c) radial velocity $\overline{u}_r$, (d-f) azimuthal velocity $\overline{u}_\theta$, and (g-i) axial velocity $\overline{u}_z$ as functions of radial position ($r$) at three different heights above the base of the plume: 0.2~m (left column), 0.4~m (middle column) and 0.6~m (right column). Results are shown in each panel for simulations with one through five levels of AMR, corresponding to effective grid resolutions of 15.6~mm (R1; thin black lines), 7.81~mm (R2; dashed red lines), 3.91~mm (R3; green dotted lines), 1.95~mm (R4; blue dash dot lines), and 0.976~mm (R5; thick black lines).}}
\end{figure}
%---------------------------------

%=================================
\subsection{Radial profiles of first- and second-order statistics}
%=================================
To gain a quantitative understanding of the effects of physical resolution on the simulation results, Figure~\ref{fig:vels} shows radial profiles of mean radial ($u_r$), azimuthal ($u_\theta$), and axial ($u_z$) velocities at three different heights above the base of the plume (i.e., 0.2~m, 0.4~m, and 0.6~m). The data at these three heights are time-averaged over the last 20~s of the simulation and then azimuthally averaged to produce radial profiles.  

Figures~\ref{fig:vels}(a-c) show that the mean radial velocity $\overline{u}_r$ requires relatively little grid resolution to converge, and simulations from R2 and finer give essentially identical results. The radial velocity is largely determined by conservation of mass and, as helium is buoyantly accelerated upwards, ambient fluid is entrained at a rate proportional to the buoyant flux of the plume. Since this buoyant flux is independent of the physical resolution of the simulation, so too should the radial velocity be independent of cell size. 

The azimuthal component of velocity $\overline{u}_\theta$ in Figures~\ref{fig:vels}(d-f) is also independent of physical resolution, and is approximately zero since there is nothing in the dynamics of an axisymmetric plume to generate a persistent azimuthal velocity. Figures~\ref{fig:vels}(d-f) show that the azimuthal component is more than an order of magnitude smaller than the radial component of velocity, and that it fluctuates about zero, with no trends or structure and no convergence in shape as a function of simulation resolution.  

In contrast to the quickly converging and small magnitude results for $\overline{u}_r$ and $\overline{u}_\theta$, Figures~\ref{fig:vels}(g-i) show that the axial velocity $\overline{u}_z$ has a much larger magnitude and varies substantially as the resolution changes. As previously discussed, the helium is accelerated upwards due to buoyancy and the plume necks inwards, creating a helium-air shear layer that leads to a Kelvin-Helmholtz instability and subsequent shedding of a toroidal vortex. During this process, ambient air is entrained into the core of the plume; the denser air is negatively buoyant relative to the surrounding helium and will travel down towards the center of the plume in a characteristic RT spike, creating a decrease in the average axial velocity $\overline{u}_z$ that is most apparent near the center of the plume in close proximity to the inlet.

In particular, Figure~\ref{fig:vels}(g) shows that, near the base of the plume, all five simulations are able to capture the peak in axial velocity that is located at about 0.18~m from the center of the plume. They all display similar profiles, despite slight differences in magnitude. As the helium and air undergo turbulent mixing higher into the domain, the peak shrinks and eventually disappears. Since the lower resolution simulations are not able to fully resolve the helium-air shear layer near the base of the plume, the development and subsequent break-down of the vortex ring is suppressed, leading to a delayed onset of turbulent mixing and a consistently larger axial velocity downstream of the inlet as compared to the higher resolution simulations. The axial velocity is relatively easy to predict using a coarse simulation near the base of the plume, but as turbulence plays a more dominant role higher into the domain, only the finest two simulations are comparable.

\begin{figure}[t!]
\includegraphics[scale=1]{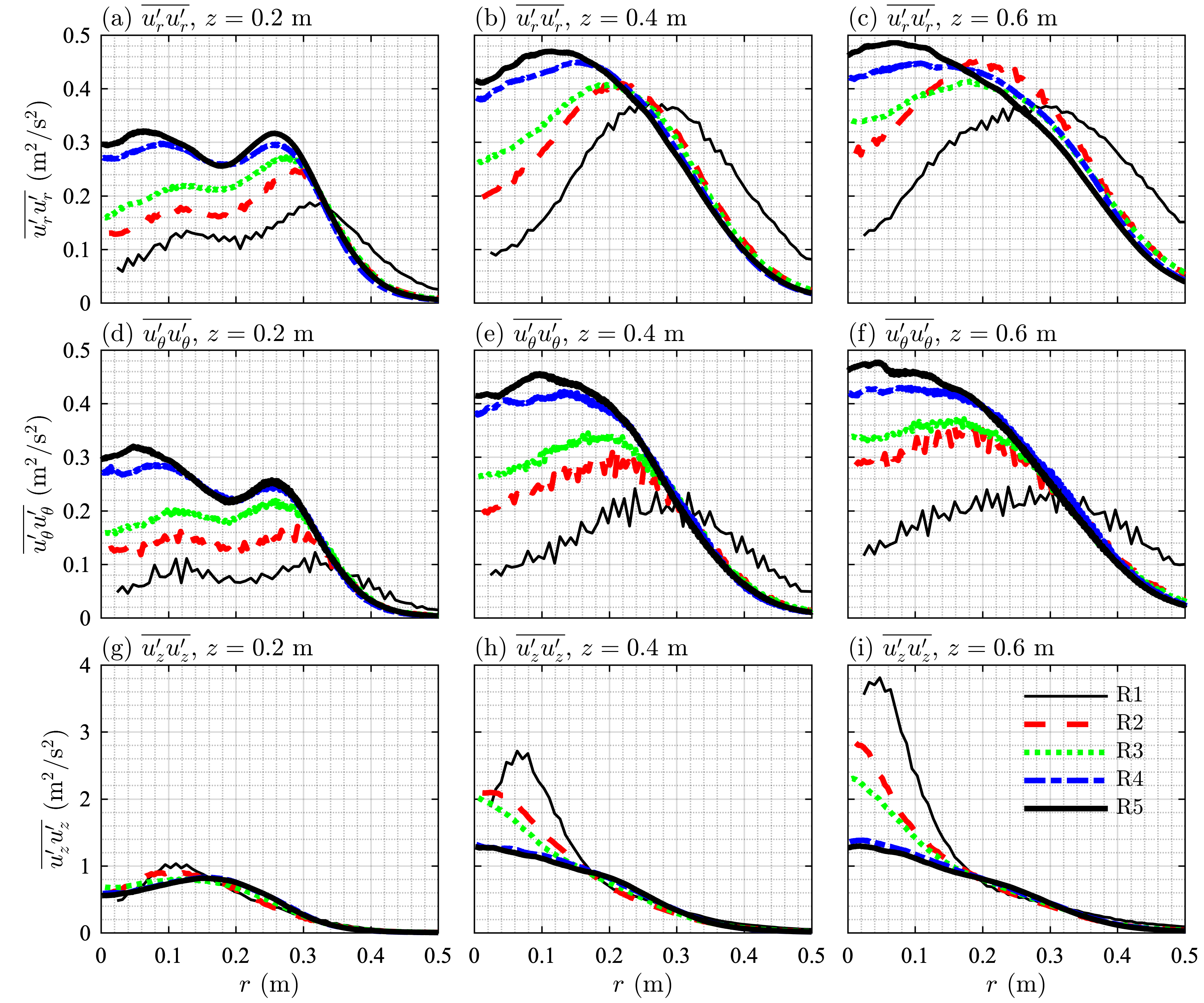}
\singlespacing{\caption{\label{fig:vars} Time and azimuthal averages of (a-c) radial velocity variance $\overline{(u'_r)^2}$, (d-f) azimuthal velocity variance $\overline{(u'_\theta)^2}$, and (g-i) axial velocity variance $\overline{(u'_z)^2}$ as functions of radial position ($r$) at three different heights above the base of the plume: 0.2~m (left column), 0.4~m (middle column) and 0.6~m (right column). Results are shown in each panel for simulations with one through five levels of AMR, corresponding to effective grid resolutions of 15.6~mm (R1; thin black lines), 7.81~mm (R2; dashed red lines), 3.91~mm (R3; green dotted lines), 1.95~mm (R4; blue dash dot lines), and 0.976~mm (R5; thick black lines).}}
\end{figure}

Consistent with the increasing importance of turbulence higher in the domain, the variances of $u_r$, $u_\theta$, and $u_z$ in Figure~\ref{fig:vars} show that the there are substantial variations as the resolution increases, but that results from the two finest resolutions (R4 and R5) are similar. The only notable deviations from total grid convergence occur in the variance of radial velocity in Figures~\ref{fig:vars}(a-c) and azimuthal velocity in Figures~\ref{fig:vars}(d-f). These deviations show that there is still some information gained when moving from 1.95~mm to 0.976~mm, although negligible additional changes are expected (for considerable additional computational cost) if moving to a further refined simulation with 0.488~mm finest resolution. 

%---------------------------------
%---------------------------------

%---------------------------------
\begin{figure}[t!]
\includegraphics[scale=1]{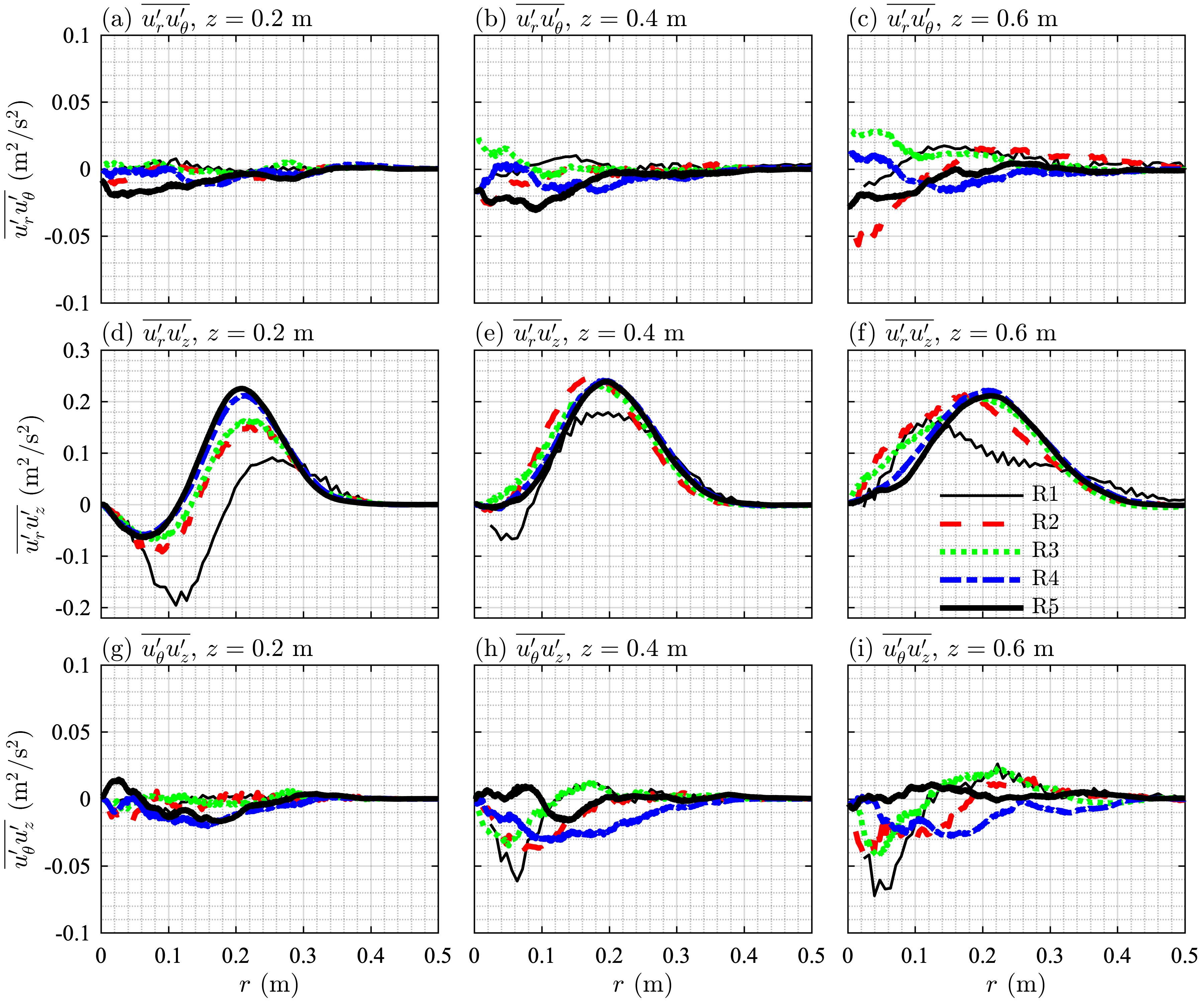}
\singlespacing{\caption{\label{fig:stresses} Time and azimuthal averages of shear stresses (a-c) $\overline{u'_r u'_\theta}$, (d-f) $\overline{u'_r u'_z}$, and (g-i) $\overline{u'_\theta u'_z}$ as functions of radial position ($r$) at three different heights above the base of the plume: 0.2~m (left column), 0.4~m (middle column) and 0.6~m (right column). Results are shown in each panel for simulations with one through five levels of AMR, corresponding to effective grid resolutions of 15.6~mm (R1; thin black lines), 7.81~mm (R2; dashed red lines), 3.91~mm (R3; green dotted lines), 1.95~mm (R4; blue dash dot lines), and 0.976~mm (R5; thick black lines).}}
\end{figure}
%---------------------------------

%---------------------------------
\begin{figure}[t!]
\includegraphics[scale=1]{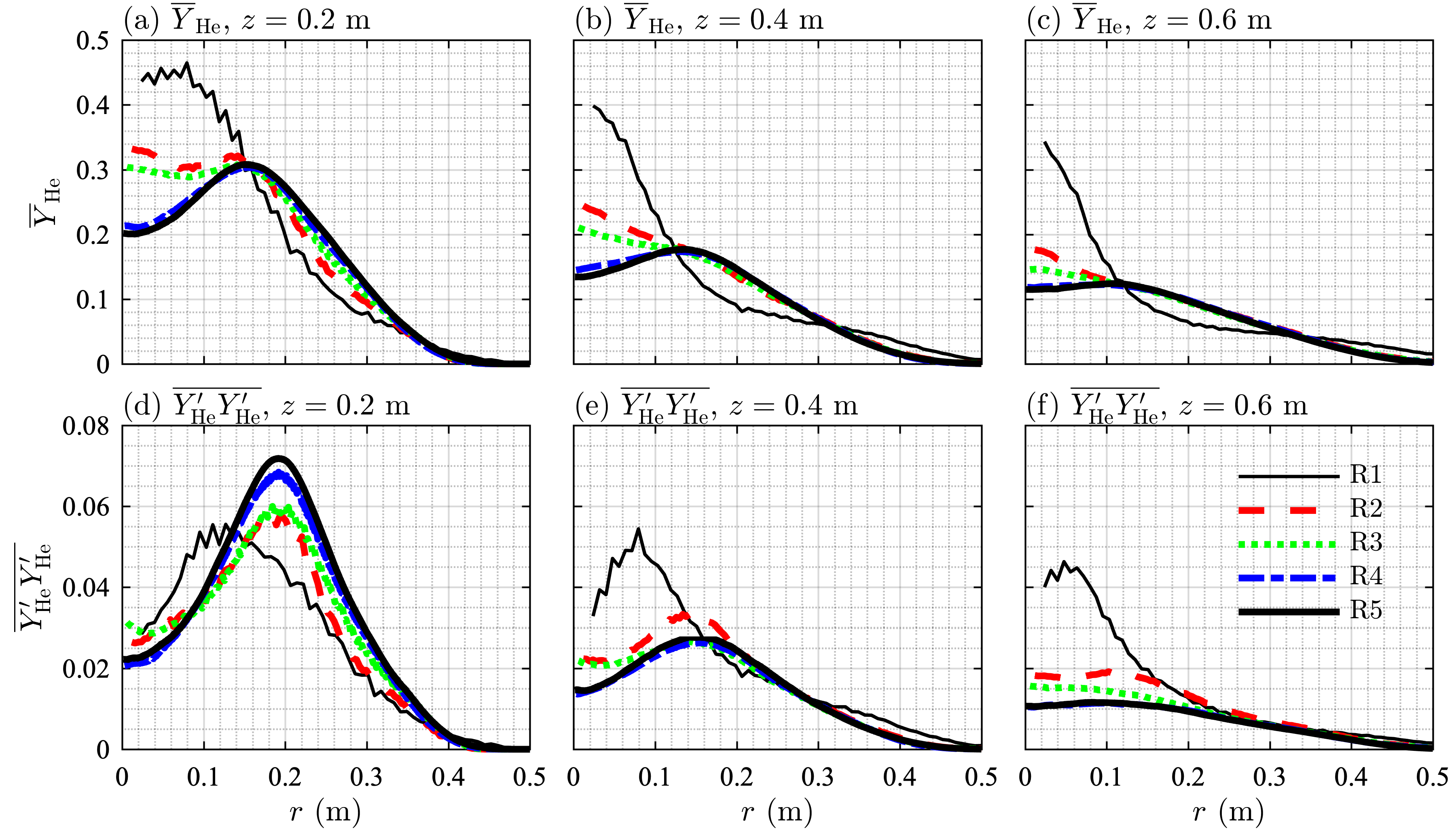}
\singlespacing{\caption{\label{fig:helium} Time and azimuthal averages of (a-c) helium mass fraction $\overline{Y}_\mathrm{He}$ and (d-f) helium mass fraction variance $\overline{(Y'_\mathrm{He})^2}$ as functions of radial position ($r$) at three different heights above the base of the plume: 0.2~m (left column), 0.4~m (middle column) and 0.6~m (right column). Results are shown in each panel for simulations with one through five levels of AMR, corresponding to effective grid resolutions of 15.6~mm (R1; thin black lines), 7.81~mm (R2; dashed red lines), 3.91~mm (R3; green dotted lines), 1.95~mm (R4; blue dash dot lines), and 0.976~mm (R5; thick black lines).}}
\end{figure}
%---------------------------------

The variances of radial and azimuthal velocities shown in Figure~\ref{fig:vars} are relatively similar in both shape and magnitude. Since the mean azimuthal velocity is effectively zero, the fluctuations in $u_\theta$ are due to turbulent fluctuations that also exist in $u_r$ and $u_z$. Figures~\ref{fig:vars}(d-f) show that the turbulent fluctuations in $u_\theta$ increase with height above the plume and are on the order of 0.5~m$^2$/s$^2$. The turbulent fluctuations in the radial component of velocity, as shown in Figure~\ref{fig:vars}(a-c), are almost identical to that of the azimuthal velocity, with a small increase in variation near the region of the developing shear instability between 0.15~m and 0.3~m.  Figures~\ref{fig:vars}(a-f) together show that, although there is no mean bulk azimuthal velocity, substantial turbulent fluctuations can lead to large variations in any given velocity component. These fluctuations tend to increase with the fidelity of the simulation since the larger cell sizes in the coarse simulations prevent the formation of large, fine-scale turbulent fluctuations.

The variance of axial velocity in Figure~\ref{fig:vars}(g) near the base of the plume exhibits a peak in the region of the shear instability, which occurs at a slightly greater radial distance from the center of the plume for the higher fidelity cases. Higher into the domain, as shown in Figures~\ref{fig:vars}(h) and (i), the variance of axial velocity for the coarser cases is greater due to the continued mixing taking place higher into the domain; the finer simulations have a relatively low variance at these heights since the plume is more well-mixed. Notably, results for R4 and R5 are nearly identical, indicating that the simulations are converged in the statistics for $u_z$ up to at least second order. 

Statistics of the remaining second order shear stresses are shown in Figure~\ref{fig:stresses}. The stresses involving $u_\theta$, shown in Figures~\ref{fig:stresses}(a-c) for $\overline{u'_r u'_\theta}$ and in Figures~\ref{fig:stresses}(g-i) for $\overline{u'_\theta u'_z}$ are close to zero at all heights above the plume due to the axisymmetric nature of the flow. The shear stress $\overline{u'_r u'_z}$ shown in Figures~\ref{fig:stresses}(d-f) is, however, much larger in magnitude and is similar for simulations R4 and R5. Near the base of the plume, Figure~\ref{fig:stresses}(d) shows that $\overline{u'_r u'_z}$ is negative close to the centerline (i.e., small $r$), indicating a flux of higher axial velocity fluid towards the centerline, and lower axial velocity fluid away from the centerline. This region of negative $\overline{u'_r u'_z}$ is lost at locations higher above the plume, as shown in Figures~\ref{fig:stresses}(e,f). At all heights, there is a pronounced positive peak in $\overline{u'_r u'_z}$ at $r\approx 0.2$~m, which marks the location of peak shear at the edge of the plume. 

Whereas profiles of mean radial and azimuthal velocities are almost identical for varying grid resolution, and profiles of mean axial velocity are similar until higher into the domain, the profiles of mean helium mass fraction in Figure~\ref{fig:helium} display a distinct shift in shape that is most readily apparent close to the base of the plume at 0.2~m, shown in Figure~\ref{fig:helium}(a). In particular, at coarse resolutions, the helium mass fraction is substantially larger than in the more finely resolved simulations and displays a peak along the centerline of the plume. As the cell size decreases, the centerline peak decreases until a physical resolution of 3.91~mm (i.e., R3) where the peak is replaced by a plateau. When the cell size is decreased further to 1.95~mm (i.e., R4), a drastic change in the mean helium mass fraction is observed. Now, instead of a centerline peak, there is an off-center peak located at approximately the same location as the peak in axial velocity (see Figure~\ref{fig:vels}). At greater distances above the base of the plume, the same structure is observed; the coarser simulations display an elevated helium mass fraction along the centerline and the finer simulations exhibit an off-center peak and corresponding dip near the centerline of the plume.

This sudden decrease in mean helium mass fraction as a function of simulation cell size is due to the development of the RT instability and subsequent bubble/spike structure. As the cell size decreases, air is able to penetrate further towards the centerline of the plume as a result of the better resolved helium-air shear layer, KH roll-up, and coherent vortex shedding. Figure~\ref{fig:helium} suggests that a critical resolution exists between 3.91~mm and 1.95~mm, at which point the shear layer is appropriately resolved and there is a substantial change in structure for the mean helium mass fraction.

The peak in the variance of helium mass fraction in Figures~\ref{fig:helium}(d-f) shows the location of the region with highest mixing between helium and air. Since the helium-air shear layer is better captured in the high resolution simulations, the variance of $Y_\mathrm{He}$ grows with increasing resolution. Once the physical resolution of the simulation drops to 1.95~mm, there is a jump in the variance of $Y_\mathrm{He}$ that indicates enhanced mixing when compared to the lower resolution simulations and results in an off-center peak at about $r = 0.2$~m.

Higher into the domain, the variance of the higher resolution simulations is smaller than that of the lower resolution cases, showing that the transition to turbulent mixing occurs much faster with a decrease in cell size. The suppression of turbulent mixing by the large cell sizes leads to a larger quantity of helium, as air is not allowed to penetrate as quickly or as deeply in to the core of the plume. This helps to preserve the potential core of the plume. At higher resolutions, the air is able to reach the center of the plume, decreasing the helium mass fraction and subsequently, the buoyant force.

%---------------------------------
\begin{figure}[t!]
\includegraphics[scale=1]{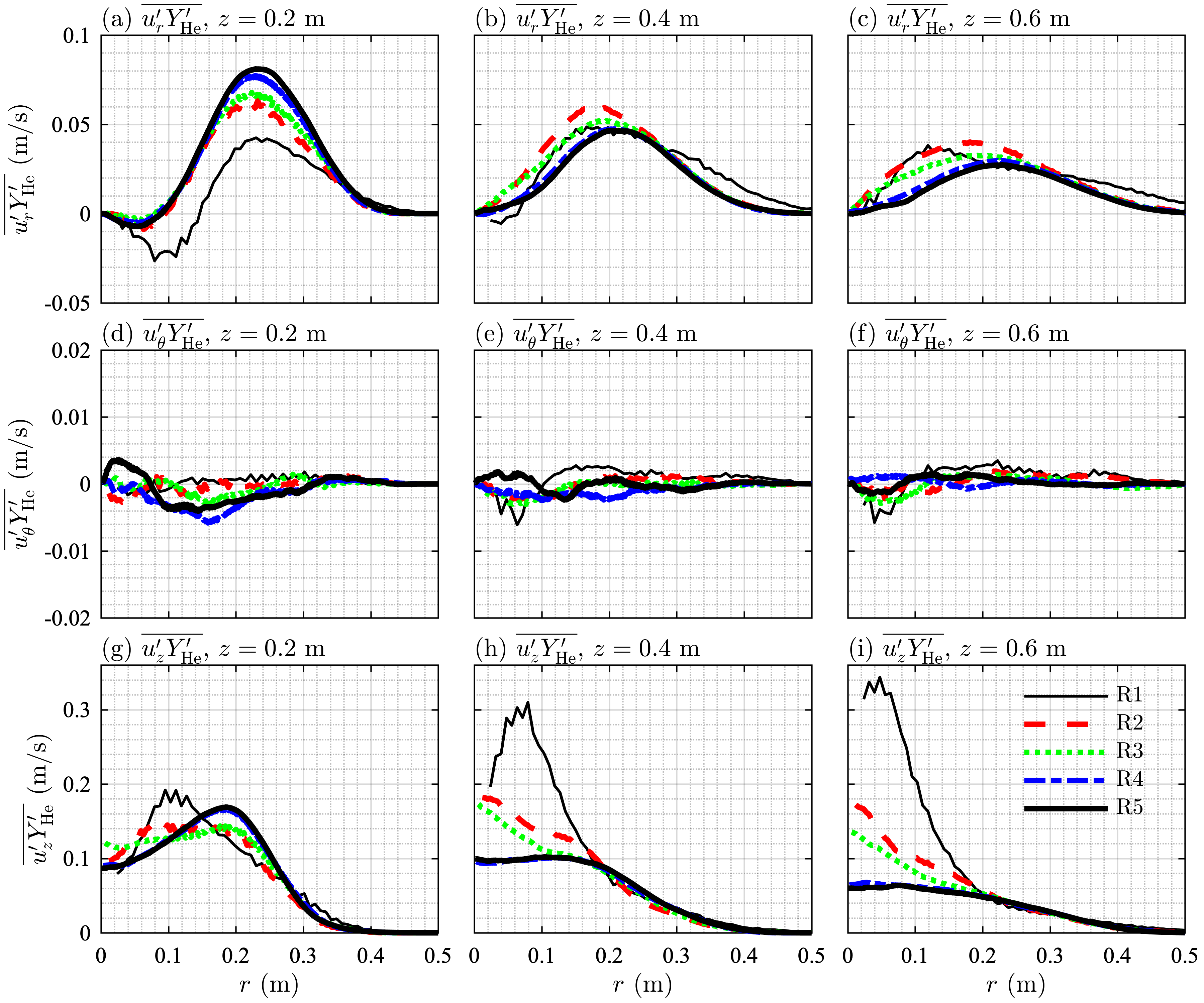}
\singlespacing{\caption{\label{fig:fluxes} Time and azimuthal averages of helium mass fraction fluxes (a-c) $\overline{u'_r Y'_\mathrm{He}}$, (d-f) $\overline{u'_\theta Y'_\mathrm{He}}$, and (g-i) $\overline{u'_z Y'_\mathrm{He}}$ as functions of radial position ($r$) at three different heights above the base of the plume: 0.2~m (left column), 0.4~m (middle column) and 0.6~m (right column). Results are shown in each panel for simulations with one through five levels of AMR, corresponding to effective grid resolutions of 15.6~mm (R1; thin black lines), 7.81~mm (R2; dashed red lines), 3.91~mm (R3; green dotted lines), 1.95~mm (R4; blue dash dot lines), and 0.976~mm (R5; thick black lines).}}
\end{figure}
%---------------------------------

Finally, Figure~\ref{fig:fluxes} shows the fluxes of helium mass fraction $Y_\mathrm{He}$ by the different components of velocity. As with the previous stress profiles in Figure~\ref{fig:stresses}, the azimuthal fluxes $\overline{u'_\theta Y'_\mathrm{He}}$ in Figures~\ref{fig:fluxes}(d-f) are close to zero due to the axisymmetry of the plume configuration. However, both $\overline{u'_r Y'_\mathrm{He}}$ in Figures~\ref{fig:fluxes}(a-c) and $\overline{u'_z Y'_\mathrm{He}}$ in Figures~\ref{fig:fluxes}(g-i) are larger in magnitude and, again, are only converged for simulations R4 and R5. Near the base of the plume, $\overline{u'_r Y'_\mathrm{He}}$ is weakly negative for small $r$, indicating that helium is transported towards the center of the plume. Further from the centerline, $\overline{u'_r Y'_\mathrm{He}}$ displays a positive peak close to $r=0.2$~m at all heights, corresponding to the mixing of helium with ambient fluid at the edge of the plume. Figures~\ref{fig:fluxes}(g-i) show that the axial flux $\overline{u'_z Y'_\mathrm{He}}$ is positive for all $r$ and all heights, although this flux is reduced for small $r$ near the base of the plume. This is due to the presence of the RT instability at the center, which brings more dense ambient air downwards towards the base of the plume along the centerline.

%=================================
\subsection{Axial profiles of first- and second-order statistics}
\begin{figure}[t!]
\centering
\includegraphics[scale=1]{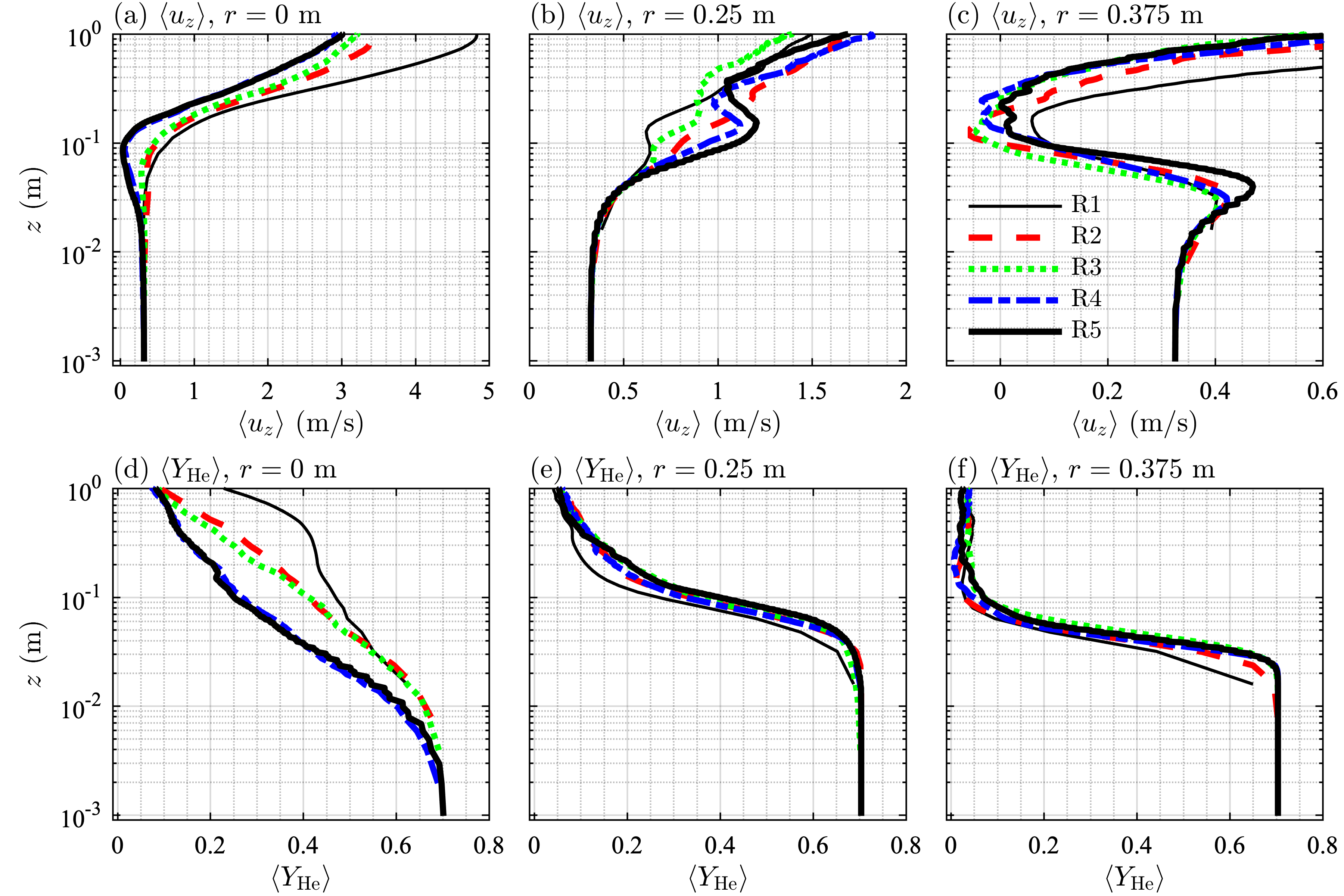}
\singlespacing{\caption{\label{fig:axial1} Time-averaged profiles of (a-c) axial velocity $\langle u_z\rangle$ and (d-f) helium mass fraction $\langle Y_\mathrm{He}\rangle$ as functions of axial position ($z$) at three different radial locations: 0~m (a,d), 0.25~m (b,e) and 0.375~m (c,f). Results are shown in each panel for simulations with one through five levels of AMR, corresponding to effective grid resolutions of 15.6~mm (R1; thin black lines), 7.81~mm (R2; dashed red lines), 3.91~mm (R3; green dotted lines), 1.95~mm (R4; blue dash dot lines), and 0.976~mm (R5; thick black lines).}}
\end{figure}

\begin{figure}[t!]
\centering
\includegraphics[scale=1]{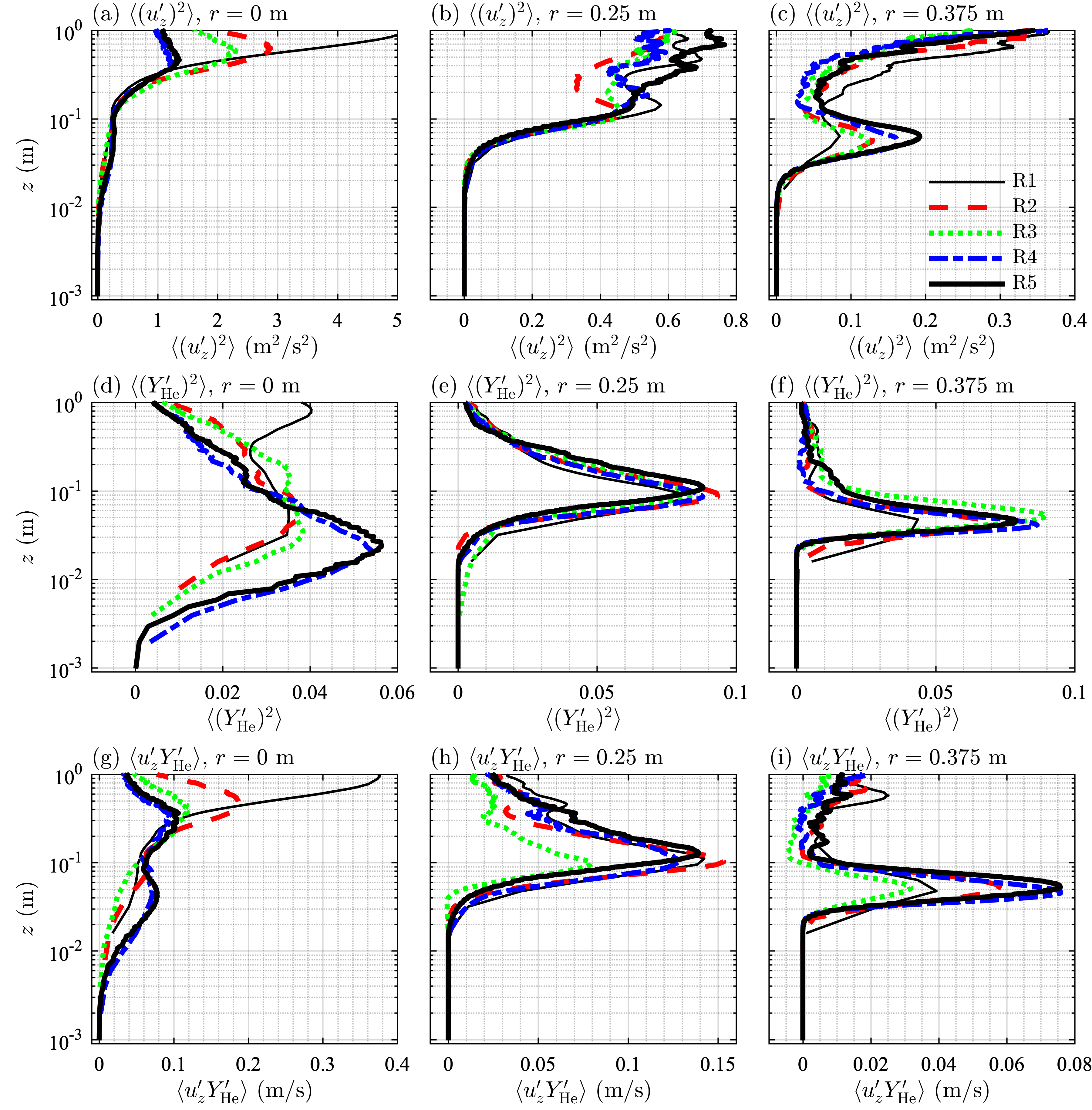}
\singlespacing{\caption{\label{fig:axial2} Time-averaged profiles of (a-c) axial velocity variance $\langle (u'_z)^2\rangle$ and (d-f) helium mass fraction variance $\langle (Y'_\mathrm{He})^2 \rangle$, and (g-i) axial flux of helium $\langle u'_z Y'_\mathrm{He} \rangle$ as functions of axial position ($z$) at three different radial locations: 0~m (left column), 0.25~m (middle column) and 0.375~m (right column). Results are shown in each panel for simulations with one through five levels of AMR, corresponding to effective grid resolutions of 15.6~mm (R1; thin black lines), 7.81~mm (R2; dashed red lines), 3.91~mm (R3; green dotted lines), 1.95~mm (R4; blue dash dot lines), and 0.976~mm (R5; thick black lines).}}
\end{figure}

%=================================
Convergence of the boundary layer at the base of the plume is indicated by the axial (i.e., vertical) profiles of first- and second-order statistics shown in Figures~\ref{fig:axial1} and \ref{fig:axial2}. These profiles are shown at three different radial locations, corresponding to the centerline of the plume, 50\% of the plume radius, and 75\% of the plume radius. The profiles of $\langle u_z\rangle$ in Figure~\ref{fig:axial1} are converged for small $z$ (i.e., roughly 1~cm and below) at each radial location for simulations R2--R5. For larger $z$, however, only the R4 and R5 simulations appear to be converged. For these two highest resolution simulations, the smallest value of $\langle u_z \rangle$ on the centerline [i.e., at $r=0$~m, shown in Figure \ref{fig:axial1}(a)] occurs at $z\approx 10$~cm, corresponding to the appearance of the RT instability that creates downward flow. Consistent with other results in this study, the coarser simulations fail to capture this RT instability.

The helium mass fraction results near the centerline in Figure~\ref{fig:axial1}(d) indicate that there is convergence for all $z$ only for R4 and R5. In particular, centerline values of $\langle Y_\mathrm{He}\rangle$ are substantially less for R4 and R5 as compared to the coarser simulations. For radial locations further from the centerline shown in Figures~\ref{fig:axial1}(e,f), $\langle Y_\mathrm{He}\rangle$ is approximately converged at all $z$ for R2--R5.

%---------------------------------
%---------------------------------

The axial profiles of second-order statistics shown in Figure~\ref{fig:axial2} are largely consistent with other results in this study. In particular, the centerline profiles of $\langle (u'_z)^2\rangle$ in Figure~\ref{fig:axial2}(a) are converged for essentially all resolutions for $z$ less than roughly $30$~cm, but only R4 and R5 are similar for larger values of $z$. This again reflects the importance of turbulent mixing higher in the domain and the need for very fine resolutions to obtain converged second-order statistics at these heights. Convergence of second-order statistics involving $Y_\mathrm{He}$, by contrast, requires very fine resolutions for all $z$, particularly on the centerline [shown in Figures~\ref{fig:axial2}(d) and (g)]. The helium mass fraction variance $\langle (Y'_\mathrm{He})^2\rangle$ in Figure~\ref{fig:axial2}(d), in particular, is poorly predicted by all but the two highest resolution simulations (i.e., R4 and R5). 

%---------------------------------
%---------------------------------

It is interesting to note that, overall, the present results indicate that there are more stringent resolution requirements for the convergence of statistics involving $Y_\mathrm{He}$, as compared to statistics involving only $u_z$. This observation may provide a partial explanation for why previous computational studies with relatively fine spatial resolutions (e.g., \cite{Chung2008,Maragkos2012,Maragkos2013}, where the resolutions were close to 1~cm) were able to obtain good agreement with the experimental results from O'Hern \emph{et al.} \cite{OHern2005} for statistics involving only $u_z$, while providing poor agreement for statistics involving $Y_\mathrm{He}$. In particular, it is possible that further improvements in grid resolution in these prior studies would have resulted in further changes in the statistics involving $Y_\mathrm{He}$, while the statistics involving $u_z$ were likely already converged.

%=================================
\subsection{Analysis of puffing frequency\label{subsec:puffing}}
%=================================
Characteristic of many buoyancy-dominated flows is the frequency at which vortical structures are shed, often referred to as the ``puffing'' frequency. For axiymmetric forced plumes, Cetegen and Kasper~\cite{Cetegen1996} performed a series of experiments to empirically determine that the puffing frequency, $f$, is related to the Richardson number of the plume, $\mathrm{Ri} = [(1-\rho_p/\rho_\infty)g D]/V_0^2$, by $f = V_0/D (0.8 \mathrm{Ri}^{0.38})$, where $\rho_\infty$ is the ambient density, $\rho_p$ is the density of the plume gas, $g$ is the magnitude of the gravitational acceleration, $D$ is the diameter of the plume, and $V_0$ is the initial velocity of the forced plume. For $\mathrm{Ri}$ above 100, this empirical relationship was found to break down and a secondary scaling took over at a higher frequency [namely, $f = V_0/D (2.1 \mathrm{Ri}^{0.28})$]. This might suggest that for flows with high Reynolds and Richardson numbers, the development of large, toroidal vortical structures from the helium-air shear layer begins to break down the empirical relationship obtained for smaller $\mathrm{Ri}$~\cite{Burton2009}.

%---------------------------------
\begin{figure}
\centering
\includegraphics[scale=1]{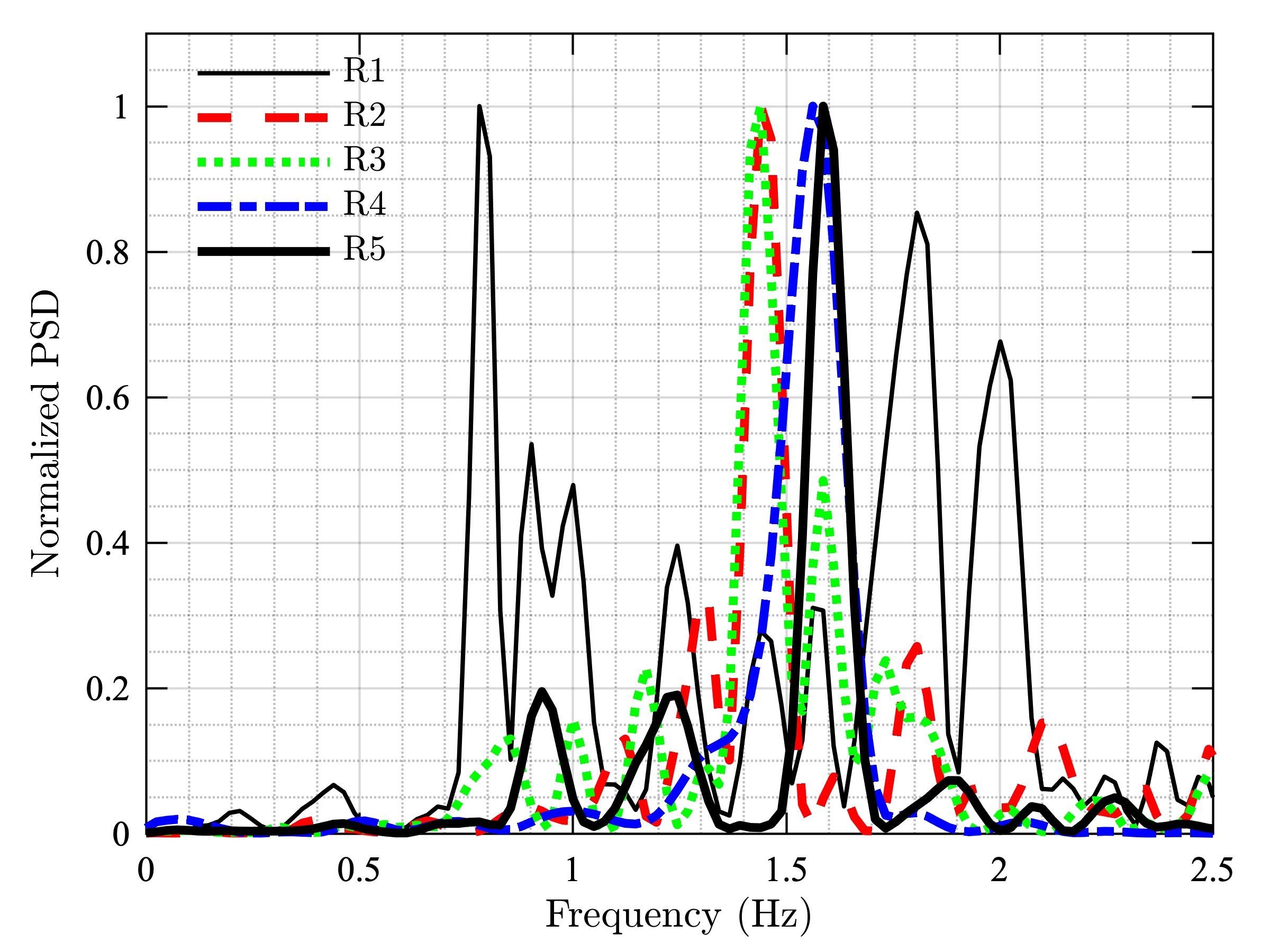}
\singlespacing{\caption{\label{fig:SVD} Frequency spectra from the singular value decomposition (SVD) performed on a time-series of $r-z$ slices taken along the center of the plume. Results are shown for simulations with one through five levels of AMR, corresponding to effective grid resolutions of 15.6~mm (R1; thin black lines), 7.81~mm (R2; dashed red lines), 3.91~mm (R3; green dotted lines), 1.95~mm (R4; blue dash dot lines), and 0.976~mm (R5; thick black lines).}}
\end{figure}
%---------------------------------

Figure~\ref{fig:SVD} shows the frequency spectra of the five different simulations obtained via singular value decomposition (SVD) of a time series of two-dimensional ($r-z$) axial velocity slices taken along the center of the plume (i.e., along $\theta=0^\circ$ and $180^\circ$). Fast Fourier Transforms are often used for this purpose, with a probe placed in the center of the plume located one-half diameter above the base of the plume. This method was satisfactory for the lower resolution simulations, but as the cell size decreased, the turbulent fluctuations along any given point in the plume polluted the spectra with high frequency content, yielding a large variety of spectra for a small change in probe position. By contrast, the SVD highlights the dominant frequencies of motion and takes into account the whole two-dimensional slice of data. Consequently, the SVD is not subjected to point-by-point variations, thus creating more consistent spectra.

The lowest resolution simulation (i.e., R1) has the most variable spectrum, with notable peaks at frequencies of roughly 0.8~Hz, 1.8~Hz, and 2.0~Hz. The presence of three or more frequencies is likely due to the coarseness of the simulation and does not represent the true puffing frequency. At a resolution of 7.81~mm (i.e., simulation R2), the dominant peak is located at roughly 1.43~Hz, and at 3.91~mm resolution (i.e., R3), there are two clear peaks, one at a frequency of roughly 1.4~Hz, and a sub-peak at a higher frequency of 1.6~Hz. At 1.95~mm resolution (i.e., R4), the peak is located at about 1.58~Hz and at a resolution of 0.976~mm (i.e., R5), the peak shifts slightly to 1.6~Hz.

Based on the empirical relation proposed by Cetegen and Kasper~\cite{Cetegen1996}, a 1~m diameter axisymmetric buoyant jet with $\mathrm{Ri}\approx 80$, as in the present case, should have a characteristic puffing frequency of about 1.4~Hz. As $\mathrm{Ri}$ increases past 100, however, the puffing frequency would be expected to shift to a higher value of 2.33~Hz based on the results in Ref.~\cite{Cetegen1996}. The present peak for R5 at 1.6~Hz thus lies between these two results, but is closer to the frequency of 1.4~Hz predicted in Ref.~\cite{Cetegen1996} using the empirical relation for $\mathrm{Ri}<100$. The observation that the puffing frequency for the present $\mathrm{Ri}\approx 80$ case is slightly greater than that predicted by Cetegen and Kasper~\cite{Cetegen1996} does not invalidate the earlier scaling law; rather, the present study merely indicates that the transition between the two scaling relations occurs slightly before $\mathrm{Ri}=100$  

From a physical standpoint, our results show that when the resolution is low, the helium mass fraction remains high near the centerline of the plume, maintaining a potential core region of the plume that is otherwise eradicated at higher resolutions due to the formation of the RT instability and the subsequent presence of a pocket of helium-air mixture. When this pocket is not allowed to form due to low resolution, we are able to match the puffing frequency predicted by the empirical relation from Ref.~\cite{Cetegen1996} for $\mathrm{Ri}<100$. When the pocket does form, the frequency of puffing shifts to a higher value, although still smaller than the secondary scaling presented in Cetegan \emph{et al.}~\cite{Cetegen1998a}. 

The helium-air shear layer and resulting instability is the primary mechanism by which air is entrained into the core of the plume. As this shear layer is better resolved, more air is allowed to penetrate the center of the plume. The air mixes with the helium, creating a region of lower buoyancy than the surrounding flow with higher helium mass fraction, forming a ``pocket'' near the center of the base of the plume. This effectively changes the structure of the flow compared to the simulations that are not capable of forming this ``pocket".

%%%%%%%%%%%%%%%%%%%%%%%%%%%%%%%%%%
\section{Conclusions and Future Work\label{sec:conclusions}}
%%%%%%%%%%%%%%%%%%%%%%%%%%%%%%%%%%
Results for five simulations of a 1~m diameter helium plume have been presented, showing that the physical resolution of the simulation has a large bearing on the mean and transient dynamics of the flow. As the resolution improves, the helium-air shear layer is better resolved until a critical resolution is reached between 3.91~mm and 1.95~mm, below which the dynamics of the plume do not change drastically. This change in dynamics can be attributed to the origin of turbulence in buoyancy driven flows that must be accurately modeled. Traditional SGS models assume a propagation of energy from large scales to smaller scales. Buoyant plumes generate turbulence via a shear layer between the plume and ambient fluid; the inverse cascade of energy from small scales to larger scales is crucial to consider when simulating buoyancy-driven flows with or without an SGS model.

For simulations that do not resolve past the critical resolution, the helium mass fraction near the center of the plume is consistently greater than in simulations that do resolve the critical resolution, consistent with prior computational studies that had a finest resolution of close to 1~cm \cite{Chung2008,Maragkos2012,Maragkos2013}. The under-resolved simulations display a potential core region of the plume, similar to that of buoyant jets, and the characteristic puffing frequency is similar to that predicted for smaller Richardson numbers in the experimental work of Cetegen and Kasper~\cite{Cetegen1996}. When the critical resolution is reached, the potential core region is replaced by a pocket of intense helium-air mixing and the puffing frequency is increased. This represents a transition to a secondary scaling for high Reynolds number plumes at a Richardson number that is close to, but slightly lower than, that observed previously~\cite{Cetegen1998a}.

Future work is required to further explore the quantitative dynamics of the plume, including the vorticity and kinetic energy dynamics. Although the present study has shown that statistics up to second order are converged for the finest resolution considered here, an additional convergence study is required for each of the physical effects (e.g., vortex stretching, kinetic energy dissipation) that determine the vorticity and kinetic energy dynamics. Additionally, more work is needed to investigate the observed transition in puffing frequency for large scale plumes. Finally, a similar resolution study for large scale pool fires should be conducted to determine whether a similar pocket of air that develops in the center of the plume likewise develops in the pool fire case. In a pool fire, this pocket of air in the center of the flow will help to feed the fire oxidizer as it continues to burn, which could have implications for large-scale fires and their spread.

Ultimately, we have shown through the series of simulations presented herein that statistics up to second order are converged. This convergence, combined with the exactly known setup of the simulations, make the resulting datasets ideal candidates for SGS model development and validation in the future. 

\acknowledgments
Helpful discussions with Drs.\ Andrew Nonaka, Alexei Poludnenko, and Chad Hoffman are gratefully acknowledged. NTW, ASM, JFG, JWD, GBR, and PEH were supported, in part, by the Strategic Environmental Research and Development Program under grant W912HQ-16-C-0026. CL was supported by the National Science Foundation Graduate Fellowship Program. NTW, CL, GBR, and PEH also acknowledge gift support from the 3M Company. Computing resources were provided by DoD HPCMP under a Frontier project award.

\end{document}